\documentclass[aps]{revtex4}
\usepackage{amsmath}
\usepackage{epsfig}
\usepackage{float}
\usepackage{dcolumn} 

\begin{document}
\title{\bf New shape-resonances in one dimension}  
\author{Zafar Ahmed$^{1}$, Shashin Pavaskar$^{2}$, Lakshmi Prakash$^{3}$}
\affiliation{$^1$Nuclear Physics Division, Bhabha Atomic Research Centre, Mumbai, 400085 India \\ $^2$National Institute of Technology, Surathkal, Mangalore, 575025, India \\ $^3$University of Texas, Austin, TX, 78705, USA}
\email{1: zahmed@barc.gov.in, 2: spshashin3@gmail.com, 3:lprakash@utexas.edu} 
\date{\today}
\begin{abstract}
Hitherto, a finitely thick barrier next to a well or a rigid wall has been considered the potential of simplest shape giving rise to resonances (metastable states) in one dimension $x \in (-\infty,\infty)$. In such a potential,
there are three real turning points at any energy below the barrier. Resonances are Gamow's (time-wise) decaying states with discrete complex energies $({\cal E}_n = E_n -i\Gamma_n/2)$. These are also spatially catastrophic states that manifest as peaks/wiggles in Wigner’s reflection time-delay at $E = \epsilon_n \approx E_n$. Here we explore potentials with simpler shapes giving rise to resonances $\--$ two-piece rising potentials having just one-turning point. We demonstrate our point by using rising exponential profile in various ways.
\end{abstract} 
\maketitle
In quantum mechanics, the bound state refers to trapping of a particle with a quantized energy (say, $E_0$) for ever in a potential well, $V(x)$. In this case, there are necessarily two real classical turning points $x_1,x_2$ such that $V(x_1)=E_0=V(x_2)$,  and the solution of Schr{\"o}dinger equation is such that $\psi(\pm \infty)=0$. Scattering states refer to the reflection and transmission of particles from the potential at any real positive energy. For the left incidence the solution of  Schr{\"o}dinger equation is made to satisfy $\psi(x\sim-\infty)=Ae^{ikx}+Be^{-ikx}$ and $ \psi(x\sim \infty)=C e^{ikx}.$ A shape resonance is a metastable state in which a particle is trapped temporarily due the very shape (Fig. 1) of a potential and then it leaks out of it. Automatic (spontaneous) emission of $\alpha$ particle from a nucleus [1] in radioactivity and field ionization [1] of an atom
when it is subject to an intense electric field are the well known examples. So far, the minimal shape of a one-dimensional or central potential for these states is required to be a finite barrier next to a wall (Fig. 1(a)) or a  well (Fig. 1(b)). Consequently, these potentials are such that at energies below the barrier height there are at  three real turning points (roots of $E=V(x))$.See Figs. 1(b,c). 

The shape-resonances are discrete complex energy states  which are determined by imposing [2] an out-going boundary condition of Gamow (1928) at the exit of the potential. If the barrier is on the left, one demands $\psi(x) \sim e^{-ikx}$  for $x\sim -\infty$ and on the other side of the well we impose $\psi(0)= 0$. This prescription of Gamow yields discrete complex energies ${\cal E}_n=E_n-i\Gamma_n/2$. The corresponding eigenstates decay time-wise but oscillate spatially with growing amplitude on the exit of the barrier. This characteristic behaviour  of shape resonances is called catastrophe.  Alternatively, by imposing $\psi(x \sim -\infty) = A e^{ikx}+ B e^{-ikx}$ one can obtain the complex reflection amplitude $r(E)=B/A$. Further, the shape resonances manifest as peaks in Wigner's time-delay, $\tau(E)$ [3]
\begin{equation}
\tau(E) = \hbar \frac{d\theta(E)}{dE}, \quad r(E)=R(E) e^{i\theta(E)}
\end{equation}
at $E \approx E_n$. To understand $\tau(E)$, let us take
one resonance situation
\begin{equation}
r(E)=\frac{E-E_0-i\Gamma_0/2}{E-E_0+i\Gamma_0/2}~ \Rightarrow ~\theta(E)= \tan^{-1}\frac{\Gamma_0/2}{E-E_0}
~\Rightarrow ~\tau(E)=\frac{\hbar \Gamma_0/2}{(E-E_))^2+\Gamma^2/4}.
\end{equation} 
So the quantal resonances manifest as a peak at $E=E_0$ in time-delay. this is much in the same way as the amplitude ${\cal A}$ of the oscillation of damped, periodically forced simple harmonic oscillator peaks at
the frequency $\omega=\omega_0$ as [4]
\begin{equation}
{\cal A}=\frac{{\cal A}_0}{(\omega^2-\omega_0^2)^2+\gamma^2},
\end{equation}
where $\omega_0$ is the frequency of the forcing term and
$\gamma$ is the coefficient of damping. The Wigner's time delay (1) is  a tool to detect a resonance both theoretically and experimentally.

The average life-time of $n^{th}$ resonant state is $\bar \tau_n=\hbar/ \Gamma_n$. However, $\bar\tau_0$ for the deepest lying resonance turns out to (approximately yet dominantly) depend up on the the penetrability (transmission probability) $T(E)$ of the free barrier (without the well). There fore phenomenon of cold emission of an electron from a metal, $\alpha$-decay from nucleus and even field ionization of an atom in the intense electric field in textbooks are  explained [1,3] in terms of barrier tunneling and the concept of discrete  complex energy eigenvalues of Gamow  is usually avoided.

The an-harmonic potential, $V_{\lambda}(x)=x^2/4+\lambda x^3$ (Fig. 1(b)), is the well known example in the research literature [5]. Some very interesting works on resonances can be seen in Ref. [6]. Gamow's decaying states are also known as Siegert states [7] and later
the phenomena  of discrete complex eigenvalues when atoms and molecules are subjected to intense electric fields is called loSurdo-Strak-effect [8]. Resonances are also discussed by studying scattering phase-shifts of the central potentials like $V(r)=V_0 \delta(r-a)$ [9]
and the rectangular barrier [2], where the regularity condition the wave function, $u(0)=0~ (\psi(r)=u(r)/r)$
acts like a rigid wall.
\begin{figure}
\centering
\includegraphics[height=4. cm, width=5. cm]{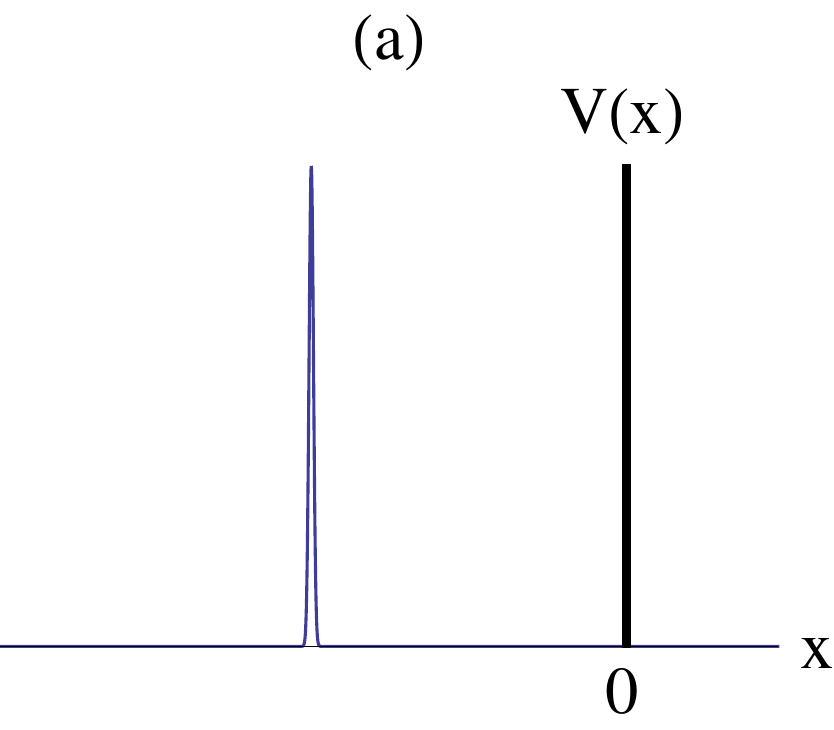}
\includegraphics[height=4. cm, width=5. cm]{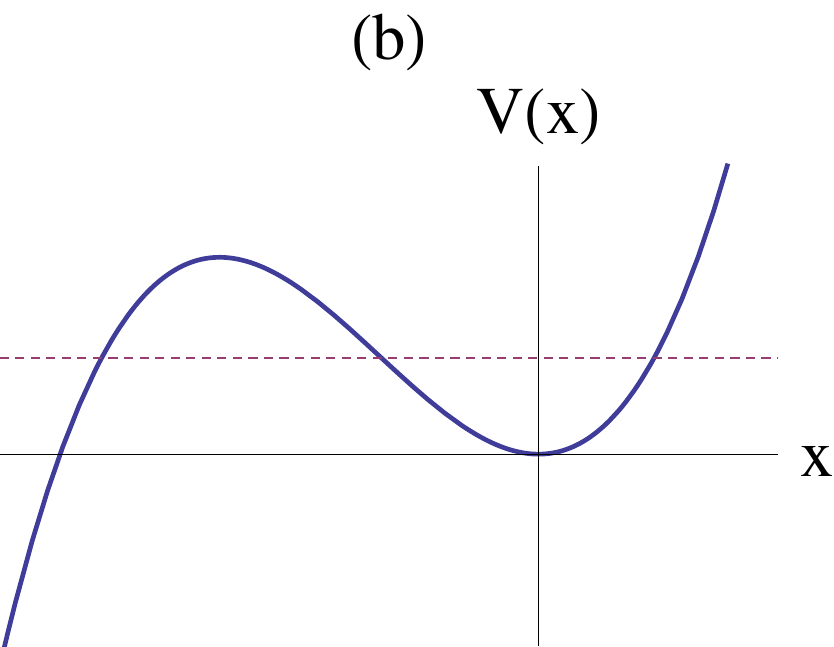}
\includegraphics[height=4. cm, width=5. cm]{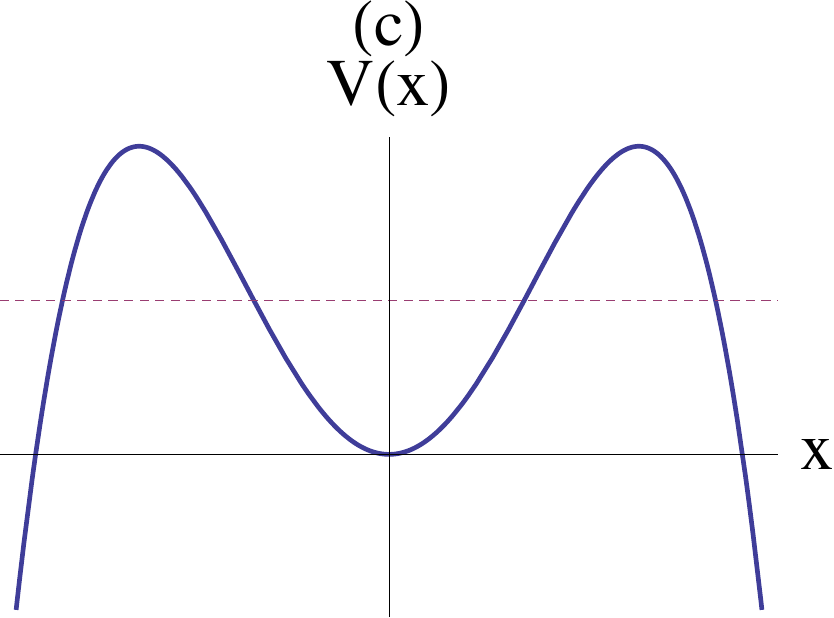}
\caption{Schematic depiction of  the simplest orthodox potentials for shape resonances. (a) Dirac delta near a rigid wall (see Eq.(3)), (b) $V_{\lambda}(x)=x^2+\lambda x^3, \lambda>0$  and (c) $V_\mu(x) = x^2-\mu x^4, \mu>0$. The horizontal dashed line cuts the potential (b) at three and (c) at four real turning points. The delta potential being extremely thin, one can also visualize three real turning points with the third one being at $x=0$, at the rigid wall itself.}
\end{figure}

However, in textbooks [3] the shape-resonances have been usually discussed by studying  a well surrounded by two side barriers ($V_\mu(x)=x^2-\mu x^4$, $\mu>0$, Fig. 1(c)) in these cases there are four real turning points turning points. Amusingly, some textbooks ask students to find the  first/second order  correction  to the bound state eigenvalues for  potentials like $V_\lambda(x)$ and $V_{\mu}(x)$, though there is no real discrete energy bound state. This happens because the discussions regarding the complex eigenvalues, ${\cal E}_n$, have been bypassed so much so that even their existence is over sighted.

Recently, the issue of scattering from a rising  potential has been initiated [10] with the study of an odd parabolic potential:$V(x\le 0)=-x^2, V(x>0)=x^2$. It turns out that for the  rising potential (say for $x>0$), one can actually demand $\psi(\infty) \sim 0$.  For $x\sim -\infty$, on the other hand, one can seek a linear combination of reflected and transmitted wave solutions of the Schr{\"o}dinger equation as per the potential for $x<0$. Unlike the usual reciprocal one-dimensional scattering, the scattering here is essentially one-sided, from left to right if $V(x)$ is rising for $x>0$ and vice-versa. One can then find the reflection amplitude $r(E)$ justifying an intuitive result that $|r(E)|^2=1$. Subsequently, one can extract the complex energy  ($E_n-i\Gamma_n/2, ~\Gamma_n > 0)$ poles of $r(E)= e^{i\theta(E)}$ (zeros of $A(E)$); they cause maxima in the Wigner's time delay $\tau(E)$. Curiously enough, the odd-parabolic potential [10] gives rise to a single peak in time-delay  and the resulting resonant state is non-catastrophic. Earlier, in an interesting analysis [11] of rectangular and Dirac delta potentials in  a semi-harmonic background,  discrete complex spectra have been found. However, this  has been unduly attributed  to the semi-harmonic (half-parabolic) potential in particular.
\begin{figure}[h]
\centering
\includegraphics[height=4. cm, width=5. cm]{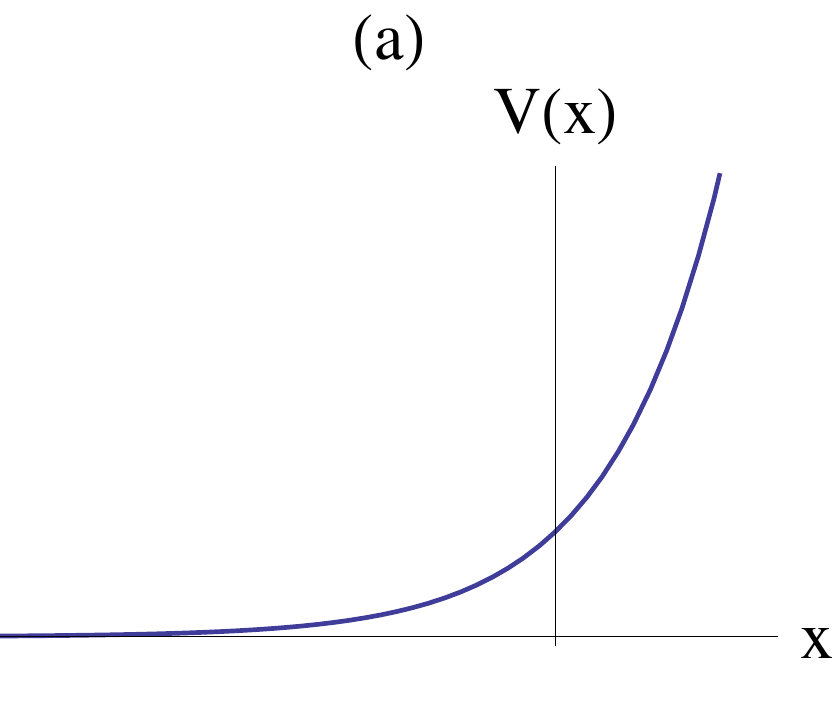}
\includegraphics[height=4. cm, width=5. cm]{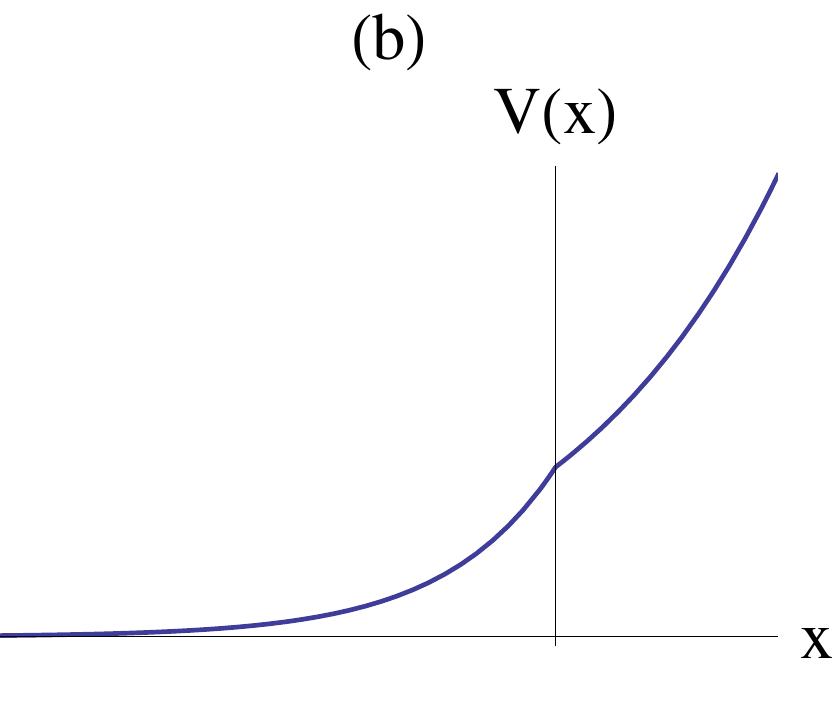}
\includegraphics[height=4. cm, width=5. cm]{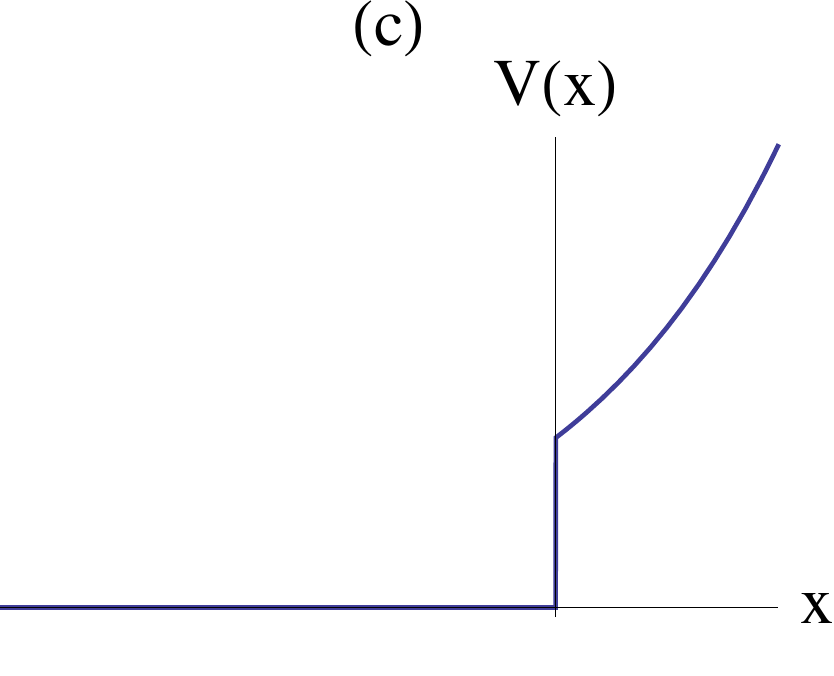}
\caption{Exponential rising potentials: (a) one-piece Eq. (9), (b) two-piece (13) when $c \ne d$, (c) when $c=0$ }
\end{figure}

In this Letter we claim that in one dimension two-piece rising potentials with only one turning point can give rise to shape resonances. 
In the following, first we demonstrate various aspects of shape 
resonances by taking the one dimensional model of Dirac Delta potential $V(x)=V_0 \delta(x+a)$ near a rigid wall at $x=0$. Next, we will discuss the one dimensional
scattering from one piece and two piece exponential rising potentials to reveal that the latter is a new model of shape resonances with only one turning point.

Historically, a particle subject to a one-dimensional potential  governed by Schr{\"o}dinger equation
\begin{equation}
\frac{d^2 \psi(x)}{dx^2}+[k^2-\frac{2m}{\hbar^2} V(x)] \psi(x)=0, \quad  k=\sqrt{\frac{2mE}{\hbar^2}}
\end{equation}
has been at the heart of many phenomena  of the micro-world and continues to be so even today. We consider
\begin{equation}
V(x)=V_0 \delta(x+a), V(x \ge 0)=\infty.
\end{equation}
for the solution of (4), which is 
\begin{equation}
\psi(x)=A e^{ikx}+ B e^{-ikx}, x<-a, \psi(x)=C \sin kx, x>-a.
\end{equation}
These two solutions should be continuous at $x=-a$. 
Due to the presence of delta function at $x=-a$, the left and right derivatives of $\psi(x)$ will mismatch at $x=-a$.
So we have
\begin{equation}
A e^{-ika}+B e^{ika}=-C \sin ka,
ik(Ae^{-ika}-B e^{ika})-C k \cos ka =V_0 C \sin ka
\end{equation}
These equations give reflection amplitude $r=\frac{B}{A}$
as
\begin{equation}
r(E)=-e^{-2ika} \frac{V_0 \sin ka + k \cos ka + i k \sin ka}{V_0 \sin ka + k \cos ka - i k \sin ka}. 
\end{equation}
Poles of $r(E)$ would imply that $A=0$ and this means that there is only Gamow's out-going wave for $x \le -a$.
Consequently, the quantized complex energies will be given by
\begin{equation}
\tan ka=\frac{k}{ik-V_0}.
\end{equation}
All roots of this equation are of the type $k_0=\alpha-i\beta$~ and corresponding energies are like ${\cal E}_0=E_0-i\Gamma_0/2$, so we can write the time dependent outgoing wave for $x \le -a$ 
\begin{equation}
\psi(x,t)= B~ e^{-ikx} e^{-iEt/\hbar} \sim B~ e^{-i\alpha x}~e^{-\beta ~ x} ~ e^{iE_0 t/\hbar} ~e^{-\Gamma_0 t/(2\hbar)}.
\end{equation}
Here $e^{-\beta ~ x}$ is the growing amplitude of the oscillatory function $e^{ikx}$ representing the spatial catastrophe in eigenstate for $x \le -a$. The factor $e^{-\Gamma_0 t/(2\hbar)}$ refers to the time-wise decay of the eigenstate. Thus, the probability of decay of such a state is given as $|\psi(x,t)|^2 \sim e^{-\Gamma_0 t/\hbar}$ characterizing it with the average decay time as $\bar\tau_0=\hbar/\Gamma_0$. Suppose there are $N_0$ number of such potentials (Fig.1(a)) having one particle each at time $t=t_0$, then after $\tau_0$ time $N_0/e$ ($e=2.71828)$ number of particle would leak out of their respective potentials. This is what happens in the case of $\alpha$ decay [1,2,] from nucleus or in the case of emission of an electron from atom when an atom is subjected to strong electric field [1,8]. 

Taking $2m=1=\hbar^2$, for $V_0=5, a=1$,
we get only one resonant state $k_0=2.7103 - 0.1779i$ as a root of (9) or the pole of (8), this gives ${\cal E}_0=7.3144-0.9648i=E_0-i\Gamma_0/2.$ This would mean that if a particle is injected at the delta barrier 
(3) with an energy $E_0=7.3144$ it will stay there for longer than at any other energy and hence it would get trapped for a finite time $\tau_0$. It would also mean that a particle will be quasi-bound
in the well between delta barrier and the rigid wall and leak out on the side $x \le -a)$. By increasing $V_0$ and $a$ we can get sharper and more number of  peaks in $\tau(E)$ giving
rise to several resonant states, here we have chosen to have a single one in Fig. 3(a). At this energy the catastrophe in eigenstate is shown in fig. 3(b) where  the scattering state, $\psi(x)$, has growing amplitude (dashed line) when $x -\infty$. The solid line represents $|\psi(x)|$ showing the spatial catastrophe rather well. This single resonance mimics the case of automatic $\alpha$-decay from nucleus. 
\begin{figure}[h]
\centering
\includegraphics[width=8 cm, height= 5 cm]{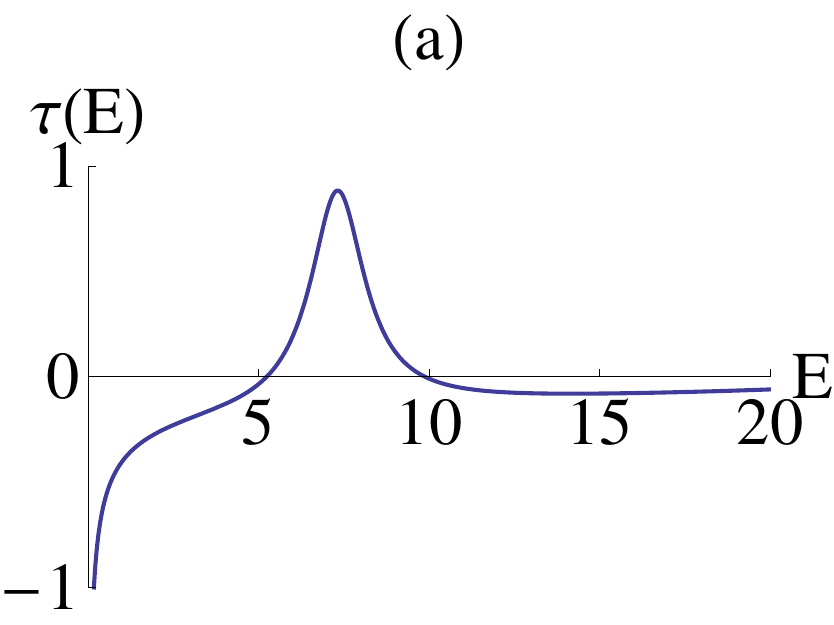}
\includegraphics[width=8 cm, height= 5 cm]{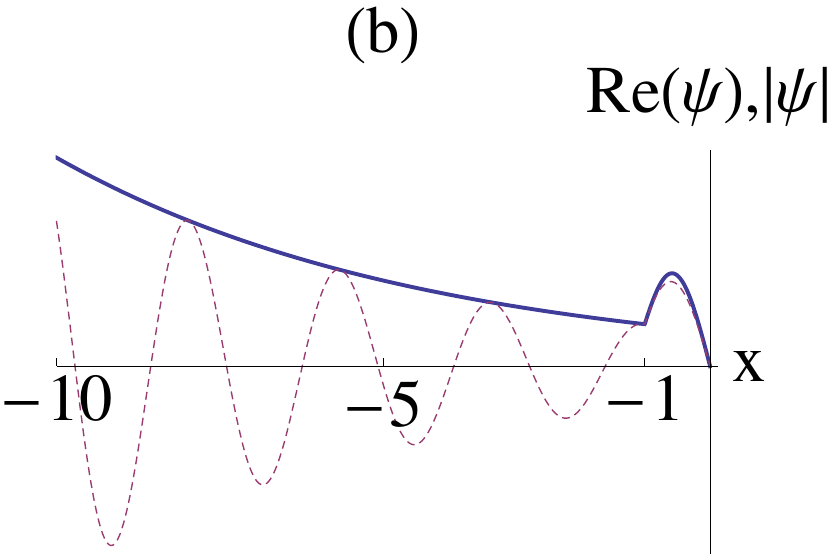}
\caption{(a):Wigner's reflection time-delay for the delta potential next to a rigid wall (Eq.3). The peak at $E=\epsilon=7.32$ in an excellent agreement with the real part of ${\cal E}=7.3144-0.9648i$ which is the pole of $r(E)$ (7). Here, we have taken $V_0=5$ and $a=1$. When the parameters are increased, we get more number of resonances sharper and multiple peaks in $\tau(E)$. (b) depiction of the spatial catastrophe in the resonant state, $\psi(x)$ at the complex energy eigenvalue ${\cal E}$. The solid line is the real part of $\psi$ and the dashed line is 
$|\psi|$. Notice the sharp corner at $x=-1$ in $|psi|$
which due to the momentum mismatch condition at $x=-a$
arsing from the presence of delta function there.}
\end{figure}
\begin{figure}
\centering
\includegraphics[width=8 cm, height= 5 cm]{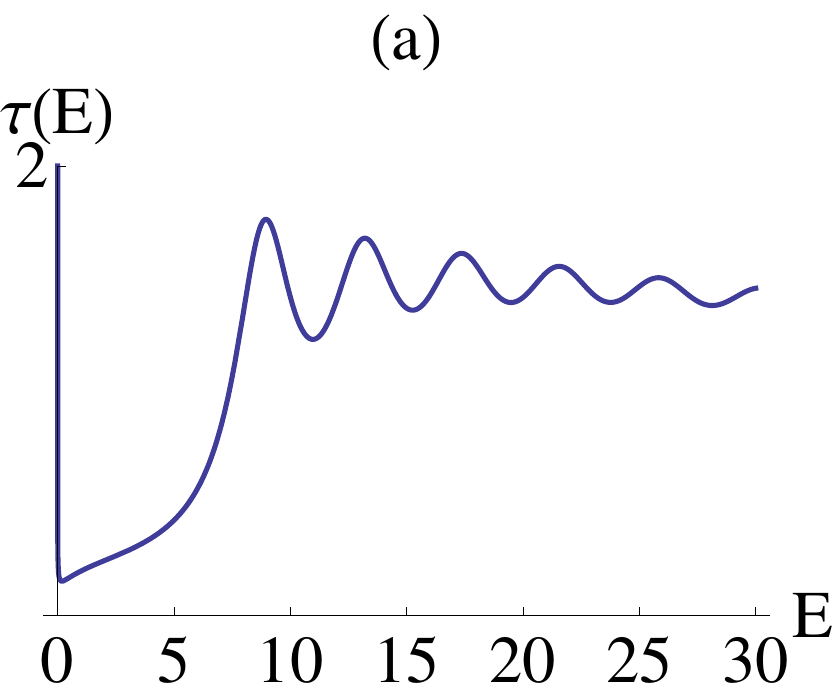}
\includegraphics[width=8 cm, height= 5 cm]{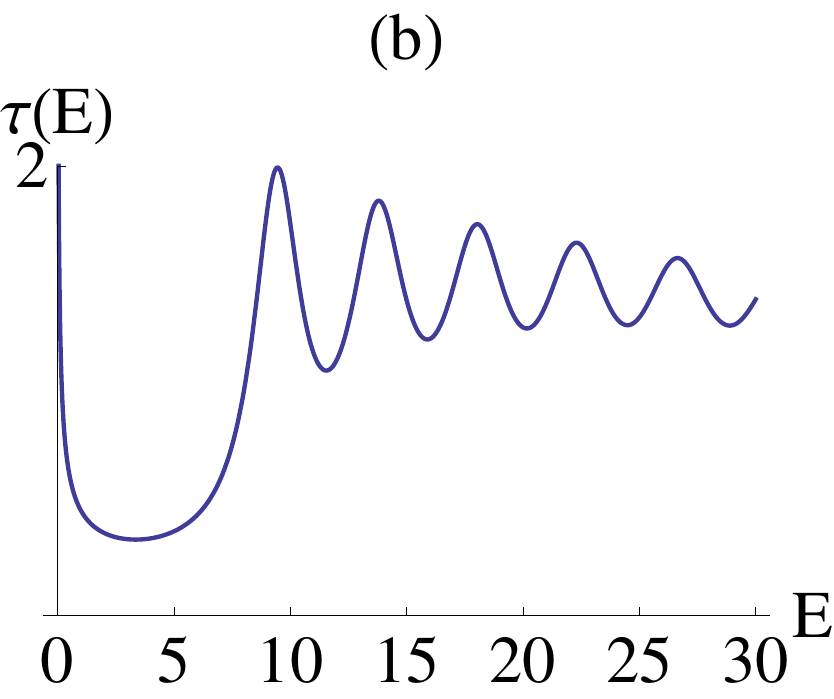}
\caption{Wigner's reflection time delay for the two-piece rising potentials in Figs. 1(b,c) is shown in (a) and (b), respectively. The parameters  $V(x)$, $E_n$, $\Gamma_n$, and $\epsilon_n$are given in Table 1. Notice an excellent agreement of $E_n$ and $\epsilon_n$ }
\end{figure}\\
Next we study the smooth one piece rising potential 
\begin{equation}
V_E(x)=V_0 e^{2x/c}, V_0, c>0.
\end{equation}

The exponential potential has been studied commonly
as an asymptotically converging central potential in the domain $(0,\infty)$, in textbooks both as a repulsive and attractive well [9]. 
The repulsive one has also been used to find complex energy eigenvalues in various ways [12]. In one dimension it provides a symmetric well or a barrier that converge asymptotically [13]. One dimensional scattering from the exponential potential, $V(x)=-V_0 e^{x/a}, V_0>0$ has been studied to find [14] the reflection and transmission coefficients. To the best of our knowledge the scattering 
from the rising potential has so far not been studied. 
The reason being that one can guess  $R(E)=1$ even intuitively. However, one can not guess the reflection phase-shift, $\theta(E)$, and hence the Wigner's reflection time-delay, $\tau(E)$. In this regard our usage
of exponential profile as (11) or (14) is new and worthwhile.
 
Let us define 
$s=\sqrt{\frac{2mV_0 c^2}{\hbar^2}},$
the Eq. (4) for (11) can be transformed to Modified cylindrical Bessel equation [15]. We seek the modified Bessel function of second kind $\psi(x)=K_{ikc}(se^{x/c})$ as the physically acceptable  solution. This vanishes for $x \sim \infty$ since [15] $K_{\nu}(z) \sim \sqrt{\frac{\pi}{2z}} e^{-z}\rightarrow 0$. Further, the identity $K_\nu(z)=\frac{I_{-\nu}(z)-I_{\nu}(z)}{\sin \nu \pi}$ [15],  and $I_{\nu}(z)\approx \frac{(z/2)^{\nu}}{\Gamma(1+\nu)}, z\sim 0$ [15], help us to write
\begin{eqnarray}
\psi(x) \sim \left\{ \begin{array}{lcr}
\sqrt{\pi/2s}~ e^{-[x/(2c)+se^{x/c}]}, \quad x\sim \infty \\
-(i\pi kc)^{-1}([(s/2)^{ikc} \Gamma(1-ikc)] e^{ikx}  \\ +[(s/2)^{-ikc} \Gamma(1+ikc)] e^{-ikx}). \quad x\sim -\infty \\
\end{array}
\right.
\end{eqnarray}
thereby giving reflection amplitude as
\begin{equation}
r(E)= -(s/2)^{-2ikc} \left (\frac {\Gamma (1+ikc)}{\Gamma(1-ikc)} \right ).
\end{equation}
It can be readily checked that the all the poles of $r(E)$ are
$ikc=-(n+1)$. These are unphysical as they give rise to a false discrete spectrum $(E_n=-(1+n)^2 \frac{\hbar^2}{2mc^2})$. Moreover, no complex energy resonances or peaks in $\tau(E)$ are  found (not shown here). In our arxiv-paper [16], we also studied two more one piece exactly solvable rising potentials, namely Morse oscillator, $V_M(x)= -V_0(2e^{x/a}-e^{2x/a}), V_0>0$, and the linear hill, $V(x)= gx$. We find that they are devoid of resonances and their $\tau(E)$ is structureless. However,
their two-piece counterparts are rich models of resonances. This demonstrate follows next. 
\begin{table*} 
\caption{First five resonances in two systems of the rising potential (14). ${\cal E}_n = E_n - i\Gamma_n/2 ~(\Gamma_n > 0)$ are the poles of $r(E)$ (16) and $\epsilon_n$ are the peak positions in time-delay, $\tau(E)$ (1). We take $2m = 1 =\hbar^2$, $V_0=5$. Notice the  closeness of $E_n$ and $\epsilon_n $, in the following cases.}
\begin{ruledtabular}[t]
\begin{tabular}{|c||c||c||c||c||c||c||c|}
		\hline
		System & $\tau(E)$. &  Parameters & ${\cal E}_0 (\epsilon_0)$ & ${\cal E}_1 (\epsilon_1)$ & ${\cal E}_2 (\epsilon_2)$ & ${\cal E}_3 (\epsilon_3)$ & ${\cal E}_4 (\epsilon_4)$\\
		\hline
		\hline
		Fig. 3(b) & Fig. 4(a)  & $c=0.5, d=5$ & $8.88 - 1.50 i$ & $13.14 - 1.87 i$ & $17.30 - 2.17 i$ & $21.51 - 2.45 i$ & $25.80 - 2.70 i$ \\
	& & &   (8.89) & (13.21) & (17.34) & (21.65) & (26.05)\\
	\hline
		Fig. 3(c) & Fig. 4(b)  &  $c = 0.0, d=5$ & $9.42 - 1.23 i$ & $13.77 - 1.49 i$ & $18.01 - 1.69 i$ & $22.28 - 1.89 i$ & $26.62 - 2.07 i$ \\
	 & & &   $(9.36)$ & $(13.46)$ & $(18.04)$ & $(22.14)$ & $(26.43)$\\
	\hline	
\end{tabular}
\end{ruledtabular}
\end{table*}

Now we propose to  make the exponential potential the exponential potential as two piece as 
\begin{equation}
V(x)= V_0 e^{2x/c}, ~ x \le 0;~  V_0 e^{2x/d}, ~ x>0.
\end{equation}
$V(x)$ is continuous but non-differentiable at $x=0$. Earlier, two-piece wells and semi-infinite (step) potentials have been found [17,18] to have a single deep minimum in reflectivity as a function of energy. This is opposed to the usual result of monotonically decreasing reflectivity. Again, we expect some striking difference in $\tau(E)$ due to its two-piece nature.

As discussed above, the left hand solution of (4) for (14)
can be expressed in terms of modified cylindrical Bessel functions.
For $x<0$, we seek
\begin{eqnarray}
\psi(x) = A (s/2)^{-ikc} \Gamma(1+ikc) I_{ikc}(s e^{x/c}) + B (s/2)^{ikc} \Gamma(1-ikc) I_{-ikc}(s e^{x/c}) x<0,  \nonumber \\
\psi(x) = C K_{ikd}(s\zeta e^{x/d}), x>0.
\end{eqnarray} 
Using
these solutions and introducing $ \zeta = d/c$, we obtain 
\begin{eqnarray}
r(E)=-(s/2)^{-2ikc}   \frac {\Gamma (1+ikc)}{\Gamma(1-ikc)}
 \left [ \frac{I_{ikc}(s) K'_{ikd}(s\zeta)-I'_{ikc}(s) K_{ikd}(s\zeta)}{I_{-ikc}(s) K'_{ikd}(s\zeta)-I'_{-ikc}(s) K_{ikd}(s\zeta)} \right].
\end{eqnarray}
When $c=d$, both the numerator and denominator in the square bracket are Wronskian functions: $[I_{\nu}(z), K_{\nu}(z)]=1/z$ [15] (real here). These cancel out, thereby giving us (13) back. But when $c\neq d$, $K_{i\nu}(z)$ is real for
real $\nu$ and $z$; the square bracket is uni-modular. Hence, it changes only the phase of $r(E)$ as compared to the phase of $r(E)$ in (13). $r(E)$ in (16) gives  rise to complex energy poles and corresponding peaks in time-delay, $\tau(E)$ (see Fig. 4 and Table 1). This can be seen as a direct consequence of making the potential two-piece.
The Table 1, presents and excellent agreement between $E_n$ (the real part of the complex pole of $r(E)$) and  the peak position $\epsilon_n$ of time-delay, $\tau(E)$.
These resonant states have been found to be catastrophic for $x\sim -\infty$ in the same manner as shown in Fig. 3(b).   

The aim of exploring the simpler shapes of potentials (fig. 2(b,c)) is not to replace the orthodox shapes (fig. 1), it is to rather add new avenues for the formation of resonances. In our arxiv-paper [16], we find that a rising potential juxtaposed to a potential well/step/barrier are the new and rich models of 
resonances, in general. For rising part of the potential we used exponential, linear and parabolic profiles. Schr{\"o}dinger being exactly solvable for them
we have the exact expression of Gamow's outgoing wave at the exit of the potential in a given model. Intriguingly the exactly solvable one piece rising potential profiles namely, the exponential potential (11) studied here and the Morse oscillator and the linear potentials studied in our arxiv-paper [16] are devoid of resonance. Hence, we attribute the occurrence of the new-shape resonances to rising and two-piece nature of these potentials.

We hope that both pedagogic and expository nature of the present work are well noted. The message coming from the present work is that if a particle
is injected at a two-piece rising potential it would get reflected but not without being delayed preferentially at some discrete positive energies. Given the intimate connection of one dimensional quantal scattering and the wave propagation [19] through a medium and the astonishing progress [20] in creating various novel synthetic mediums it is not surprising that the these proposed new shape resonances would be experimentally verified and harnessed further. 
\section*{Acknowledgement} We would like to thank Dr. V. M. Datar for his support and interest in this work.

\end{document}